\documentclass[twocolumn,showpacs,preprintnumbers,showkeys,superscriptaddress]{revtex4}
\usepackage{mathrsfs}
\usepackage{amssymb}
\usepackage{graphicx}
\usepackage{dcolumn}
\usepackage{bm}
\usepackage{graphicx}
\usepackage{float}

\newfont{\largemi}{cmmi10}
\newfont{\smallmi}{cmmi6}

\draft

\def\eqref#1{Eq.~(\ref{eq:#1})}

\begin{document}

\title{Residual proton-neutron interactions and the $N_{\rm p}
N_{\rm n}$ scheme }

\author{G. J. Fu}
\affiliation{Department of Physics,  Shanghai Jiao Tong University,
Shanghai 200240, China}

\author{Hui Jiang} \affiliation{Department of Physics,
Shanghai Jiao Tong University, Shanghai 200240, China}\affiliation{School of
Arts and Science, Shanghai Maritime University, Shanghai 200135, China}

\author{Y. M. Zhao \footnote{Corresponding author}}
\email{ymzhao@sjtu.edu.cn} \affiliation{Department of Physics,
Shanghai Jiao Tong University, Shanghai 200240, China}
\affiliation{Center of Theoretical Nuclear Physics, National
Laboratory of Heavy Ion Accelerator, Lanzhou 730000, China}
\affiliation{CCAST, World Laboratory, P.O. Box 8730, Beijing 100080,
China}

\author{A. Arima}
\affiliation{Department of Physics, Shanghai Jiao Tong University,
Shanghai 200240, China} \affiliation{Science Museum, Japan Science
Foundation, 2-1 Kitanomaru-koen, Chiyoda-ku, Tokyo 102-0091, Japan}
\date{\today}

\date{\today}

\begin{abstract}
We investigate the correlation between integrated proton-neutron
interactions obtained by using the up-to-date experimental data of
binding energies and  the $N_{\rm p} N_{\rm n}$, the product of
valence proton number and valence neutron number with respect to the
nearest doubly closed nucleus. We make corrections on a previously
suggested formula for the integrated proton-neutron interaction. Our
results demonstrate a nice, nearly linear, correlation between the
integrated p-n interaction and $N_{\rm p} N_{\rm n}$, which provides
us with a firm foundation of the applicability of the $N_{\rm p}
N_{\rm n}$ scheme to nuclei far from the stability line.

\end{abstract}


\vspace{0.4in}

\maketitle

\newpage

The importance of the residual proton-neutron   interaction in the
nuclear structure has long been emphasized \cite{1,2,17} in
evolution of single-particle structure, the development of
collectivity, phase transition and deformation.  Based on this
scenario, Casten suggested \cite{3,5} that a simple quantity,
$N_{\rm p}N_{\rm n}$, the product of valence proton number $N_{\rm
p}$ and valence neutron number $N_{\rm n}$with respect to the
nearest doubly closed nucleus, is a reasonable measure for the
strength of the residual proton-neutron interaction, and that this
simple quantity is both interpretive in classification of collective
motion for low-lying states and predicative in studying unknown
regions. See Ref. \cite{16} for a comprehensive review. Recent
developments and applications along this line can be found in refs.
\cite{8,23,24}.

In ref. \cite{10} Zhang $et$ $al.$ suggested a very simple approach
to extract the proton-neutron interaction of the last proton with
the last neutron ($\delta$$V_{\rm pn}$) for odd-odd nuclei by
experimental data of binding energies of their even-even neighboring
nuclei, via a double difference procedure.  This approach attracted
much attention since then \cite{15,23,27,28,29,30,31}. In the same
paper Zhang {\it et al.} also suggested a formula of the integrated
proton-neutron interaction ($V_{\rm pn}$). They plotted the $V_{\rm
pn}$ versus $N_{\rm p}N_{\rm n}$ for nuclei with mass number around
100 and 130.  Their results showed that $V_{\rm pn}$ is
approximately proportional to the value of $N_{\rm p}N_{\rm n}$,
with deviations for small $N_{\rm p}N_{\rm n}$.

Ref. \cite{10}  was published more than two decades ago. At that
time experimental data of binding energies to perform systematic
investigation of this formula were restricted to a limited number of
nuclei. In the last two decades piles of new experimental data of
binding energies became available \cite{12}, due to the new
radioactive beam facilities worldwide. It would be interesting to
revisit the integrated proton-neutron interactions and to
investigate whether or not there exist similar relations between
$V_{\rm pn}$ and $N_{\rm p}N_{\rm n}$ for unstable nuclei. In doing
so, we suggest other formulas to extract the integrated
proton-neutron interactions.

Let us begin with the famous formula of the residual proton-neutron
interaction of the last proton with the last neutron
($\delta$$V_{\rm pn}$) for odd-odd nuclei suggested in ref.
\cite{10},
\begin{eqnarray}
  && \delta V_{\rm pn}(Z, N) \nonumber \\
  &&=\frac{1}{4}\{[B(Z+1, N+1)-B(Z+1, N-1)] \nonumber \\
  &&-[B(Z-1, N+1)-B(Z-1, N-1)]\},  \label{Zhang1}
\end{eqnarray}
where $B$ is the nuclear binding energy, both proton number $Z$ and
neutron number $N$ are odd.

By summing $\delta$$V_{\rm pn}$ over all the valence protons and
valence neutrons, the  integrated proton-neutron interaction $V_{\rm
pn}$ for an odd-odd nucleus $(Z+\delta_{\rm p}, N+\delta_{\rm n})$
was obtained by Zhang {\it et al.} in Ref. \cite{10}. Their formula
is given as follows.
\begin{eqnarray}
  &&V_{\rm pn}(Z+\delta_{\rm p}, N+\delta_{\rm n}) \nonumber \\
  &&=\delta_{\rm p}\delta_{\rm n}\cdot\sum_{Z_{\rm x}
  =\frac{Z_{0}}{2}}^{\frac{Z}{2}}\sum_{N_{\rm x}
  =\frac{N_{0}}{2}}^{\frac{N}{2}}4\cdot\delta V_{\rm pn}(2Z_{\rm x}+\delta_{\rm p}, 2N_{\rm x}+\delta_{\rm n})\nonumber \\
  &&=\delta_{\rm p}\delta_{\rm n}\{[B(Z+2\delta_{\rm p}, N+2\delta_{\rm n})-B(Z+2\delta_{\rm p}, N_{0})] \nonumber \\
  &&-[B(Z_{0}, N+2\delta_{\rm n})-B(Z_{0}, N_{0})]\}, \label{Zhang2}
\end{eqnarray}
where $Z_{0}$ and $N_{0}$ are the nearest magic numbers,
$\delta_{\rm p}$ ($\delta_{\rm n}$) is $+$1 if the valence protons
(neutrons) are particle-like and $-$1 if the valence protons
(neutrons) are hole-like, both $Z$ and $N$ are even numbers.

\begin{figure}
\includegraphics[width = 0.461\textwidth]{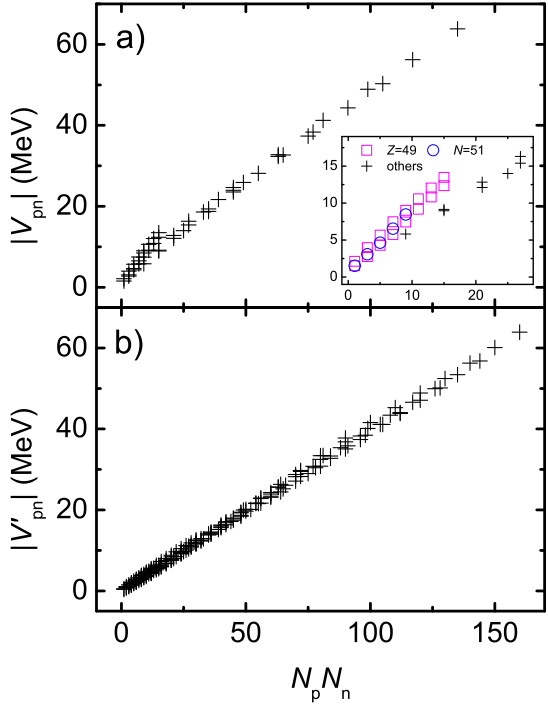}
\caption{\label{fig1} The integrated proton-neutron interaction
versus $N_{\rm p}N_{\rm n}$ for the case of 28$<$$Z$$<$50 and
50$<$$N$$<$82. (a). $|V_{\rm pn}|$  are obtained by using eq.
(\ref{Zhang2}). In panel (a) only odd-odd nuclei are included. The
insert here highlights deviations for $Z$=51 isotopes and $N$=81
isotones. (b). $|V'_{\rm pn}|$ are calculated by using our new
formula, eq. (\ref{Fu-1}). In panel (b) not only odd-odd nuclei but
also even-even and odd-mass nuclei are included. The magnitudes of
the integrated proton-neutron interaction calculated by using eq.
(\ref{Fu-1}) is slightly smaller than those obtained by using the
previous formula suggested in ref. \cite{10}. One sees that the
linear correlation between the proton-neutron interactions and
$N_{\rm p}N_{\rm n}$  is not very good in panel (a) but remains to
be good in panel (b). }
\end{figure}

\begin{figure}
\includegraphics[width = 0.461\textwidth]{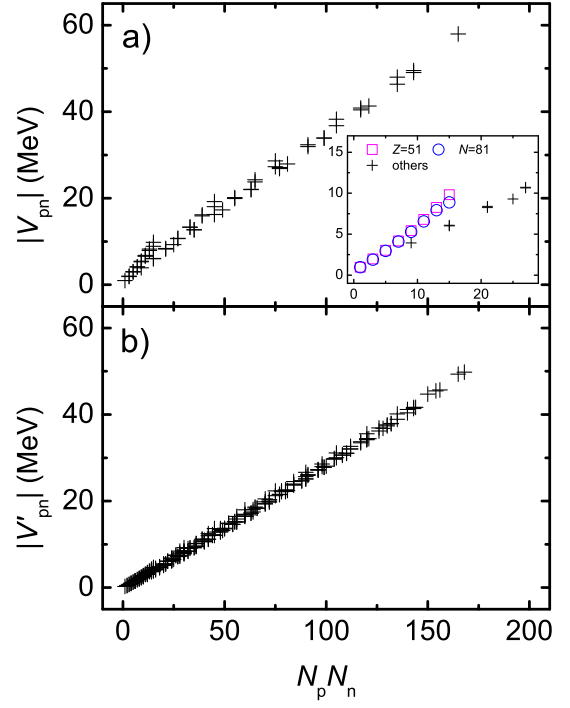}
\caption{\label{fig2} The same as Fig. 1 except 50$<$$Z$$<$82 and
50$<$$N$$<$82.  Panel (a) presents the $|V_{\rm pn}|$ calculated by
using eq. (\ref{Zhang2}), and panel (b) shows the $|V'_{\rm pn}|$
calculated by using eq. (\ref{Fu-1}). The linearity is not very good
(in particular when $N_{\rm p}N_{\rm n}$ is small) in panel (a) but
remains to be good in panel (b). }
\end{figure}

\begin{figure}
\includegraphics[width = 0.461\textwidth]{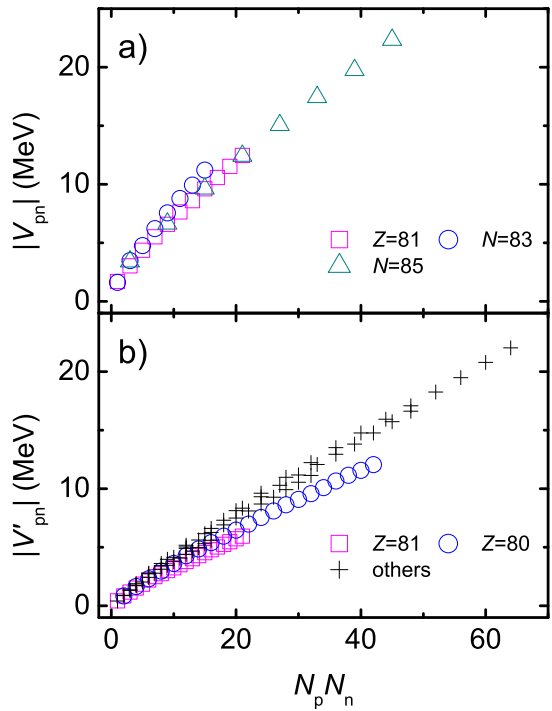}
\caption{\label{fig2} The same as Fig. 1 except 50$<$$Z$$<$82 and
82$<$$N$$<$126. Panel (a) presents the $|V_{\rm pn}|$ calculated by
using eq. (\ref{Zhang2}), and panel (b) shows the $|V'_{\rm pn}|$
calculated by using eq. (\ref{Fu-1}). }
\end{figure}

The results of $V_{\rm pn}$ by applying eq. (\ref{Zhang2}) to
odd-odd nuclei are shown in panels (a) in Figs. 1-3 (a), for nuclei
with 28$<$$Z$$<$50 and 50$<$$N$$<$82, those with 50$<$$Z$$<$82 and
50$<$$N$$<$82, and those with 50$<$$Z$$<$82 and 82$<$$N$$<$126,
respectively. The experimental data of binding energies are taken
from Ref. \cite{12} (i.e., here up-to-date experimental data of
binding energies are included in comparison to Ref. \cite{10}).
These results of $V_{\rm pn}$ are consistent with those shown in
Ref. \cite{10}: $V_{\rm pn}$ is proportional to $N_{\rm p}N_{\rm n}$
for large $N_{\rm p}N_{\rm n}$, and there are deviations from the
linear correlation for small $N_{\rm p}N_{\rm n}$. The  inserts of
Figs. 1-2 (a) highlight the details of such deviations.

We would like to point out that there are inadequacies in the
construction of the formula for Vpn (see eq. (2) above) in ref.
[10], which can be important for nuclei with smaller numbers of
valence nucleons. To see this, let us apply eq. (\ref{Zhang2}) to
the $^{128}$I nucleus.
\begin{eqnarray}
  &&V_{\rm pn}(^{128}\rm I)
   =V_{\rm pn}(53, 75)
   =V_{\rm pn}(52+1, 76-1) \nonumber \\
  &=&-\{[B(52+2, 76-2)-B(52+2, 82)] \nonumber \\
  && -[B(50, 76-2)-B(50, 82)]\} \nonumber \\
  & =&B(54, 82)-B(54, 74)   -B(50, 82)+B(50, 74)  . \label{example}
\end{eqnarray}
One sees that the above $V_{\rm pn}$ for $^{128}$I actually
corresponds to even-even nucleus $^{128}$Xe which has four valence
protons and eight valence holes with respect to $^{132}$Sn. Assuming
the proton-neutron interaction is equal for all valence particles,
this $V_{\rm pn}$ equals $32 \times \delta V_{\rm np}$, while
$^{128}$I has three valence protons and seven valence neutrons and
its $V_{\rm pn}$ should equal $21  \times \delta V_{\rm np}$.

In order to correct the above inadequacies with eq (2), i.e., for
example, $V_{\rm pn}$ for $^{128}$I approximately equals $21 \times
\delta V_{\rm np}$ but eq. (\ref{Zhang2}) gives $32 \times \delta
V_{\rm np}$,  we suggest the following formula for the integrated
proton-neutron interaction, denoted by $V'_{\rm pn}(Z, N) $. Let us
define
\begin{eqnarray}
  S(Z, N) &=& B(Z+\delta_{\rm p}, N+\delta_{\rm n})+B(Z+\delta_{\rm p}, N) \nonumber \\
   & + & B(Z, N+\delta_{\rm n})+B(Z, N) .
\end{eqnarray}
We obtain
\begin{eqnarray}
  &&V'_{\rm pn}(Z, N) \nonumber \\
  &&=\delta_{\rm p}\delta_{\rm n}\cdot\sum_{Z_{\rm x}=Z_{0}+1}^{Z}\sum_{N_{\rm x}=N_{0}+1}^{N}\delta V_{\rm pn}(Z_{\rm x}, N_{\rm x}) \nonumber \\
  &&=\frac{1}{4}\delta_{\rm p}\delta_{\rm n} \left[
   S(Z, N) + S(Z_0, N_0) \right. \nonumber \\
   &&  ~~~~~~~~~~ \left. - S(Z_0,N) - S(Z, N_0) \right] ~.
   \label{Fu-1}
\end{eqnarray}
The philosophy to construct the above formula is the same as that of
Ref. \cite{10}.  Eq. (\ref{Fu-1}) fixes the problem  arising in eq.
(\ref{Zhang2}), the problem of which is exemplified in eq.
(\ref{example}). We also note that eq. (\ref{Fu-1}) works not only
for odd-odd but also for even-even and odd-mass nuclei. For nuclei
with one valence proton (or one proton hole) and one valence neutron
(or one neutron hole) with respect to the doubly magic core, eq.
(\ref{Fu-1}) is reduced to eq. (\ref{Zhang1}).

Now let us look at $|V'_{\rm pn}|$ suggested eq. (\ref{Fu-1}) in
this paper. Here not only odd-odd but also even-even and odd-mass
nuclei are taken into account. The results of $|V'_{\rm pn}|$  for
nuclei with 28$<$$Z$$<$50 and 50$<$$N$$<$82, and for those with
50$<$$Z$$<$82 and 50$<$$N$$<$82, are shown in Figs. 1-2 (b),
respectively. Very interestingly, one sees that the deviations from
the linear $V_{\rm np}$-$N_{\rm p}N_{\rm n}$ correlation  in Figs.
1-2 (a) does not arise in the $V'_{\rm np}$-$N_{\rm p}N_{\rm n}$
plot in Figs. 1-2 (b), after our refinements.

Similarly, we show in Fig. 3 (b) the results of  $|V'_{\rm pn}|$
versus $N_{\rm p}N_{\rm n}$  for the case of 50$<$$Z$$<$82,
82$<$$N$$<$126.  One sees that $|V'_{\rm pn}|$ versus $N_{\rm
p}N_{\rm n}$ remains to be good, except slight deviations from the
linear correlation arise for nuclei with proton number or neutron
number close to the nearest magic number. For these ``anomalous"
nuclei (i.e., Z=80 and 81 isotopes, see Fig. 3(b) for details) in
Fig. 3(b), the values of $V'_{\rm pn}$ are slightly smaller than
those of ``normal" nuclei.

To summarize, in this paper we examine the linear correlation
between the integrated proton-neutron interaction obtained by using
the up-to-date experimental data of binding energies for relevant
nuclei and the $N_{\rm p}N_{\rm n}$, the product of valence proton
number $N_{\rm p}$ and valence neutron number $N_{\rm n}$with
respect to the nearest doubly closed nucleus. We suggest refinements
to the formula of the proton-neutron interaction suggested in ref.
\cite{10}. The new formula works not only for odd-odd nuclei but
also for even-even and odd-mass nuclei.

The proton-neutron interactions obtained in the present paper
exhibit excellent linear correlations in terms of $N_{\rm p}N_{\rm
n}$ for nuclei with 28$<$$Z$$<$50, 50$<$$N$$<$82, and those with
50$<$$Z$$<$82, 50$<$$N$$<$82. For nuclei with 50$<$$Z$$<$82 and
82$<$$N$$<$126, the linearity remains to be good in general,
although proton-neutron interactions  for $Z=80$ and 81 exhibit very
slight deviations.

Thus our results provide us with a firm foundation for the
applicability of the $N_{\rm p}N_{\rm n}$ scheme to nuclei far from
the stability line.

{\bf Acknowledgements:} We thank the National Natural Science
Foundation of China for supporting this work under grant 10975096.
This work is also supported partly by Chinese Major State Basic
Research Developing Program under Grant 2007CB815000, and Science \&
Technology Program of Shanghai Maritime University under grant No.
20100086.


\begin{thebibliography}{30}

\bibitem{1} A. de Shalit and M. Goldhaber, Phys. Rev. {\bf 92}, 1211 (1953).

\bibitem{2} I. Talmi, Rev. Mod. Phys. {\bf 34}, 704 (1962).

\bibitem{17} P. Federman and S. Pittel, Phys. Lett. {\bf B69}, 385
(1977);  {\bf B77}, 29 (1977);   Phys. Rev. {\bf C20}, 820.


\bibitem{3} R. F. Casten, Nucl. Phys. {\bf A443}, 1 (1985).

\bibitem{5} R. F. Casten, Phys. Rev. Lett. {\bf 54}, 1991 (1985).

\bibitem{16} For a review, see R. F. Casten and N. V. Zamfir, J. Phys. {\bf G22}, 1521 (1996).

\bibitem{8} Y. M. Zhao, R. F. Casten and A. Arima, Phys. Rev. Lett. {\bf 85}, 720 (2000).

\bibitem{23} R. B. Cakirli and R. F. Casten, Phys. Rev. Lett. {\bf 96}, 132501 (2006).

\bibitem{24} Jin-Hee Yoon, Eunja Ha and Dongwoo Cha, J. Phys. {\bf G34}, 2545 (2007);
 Guanghao Jin, Jin-Hee Yoon and Dongwoo Cha, J. Phys. {\bf G35}, 035105
 (2008).

\bibitem{10} J. Y. Zhang, R. F. Casten and D. S. Brenner, Phys. Lett. {\bf B227}, 1 (1989).

\bibitem{15} D. S. Brenner, C. Wesselborg, R. F. Casten, D. D. Warner and J. Y. Zhang, Phys. Lett. {\bf B243}, 1 (1990).

\bibitem{27} P. Van Isacker, D. D. Warner and D. S. Brenner, Phys. Rev. Lett. {\bf 74}, 4607 (1995).

\bibitem{28} R. B. Cakirli, D. S. Brenner, R. F. Casten and E. A. Millman, Phys. Rev. Lett. {\bf 94}, 092501 (2005).

\bibitem{29} D. S. Brenner, R. B. Cakirli and R. F. Casten, Phys. Rev. {\bf C73}, 034315 (2006).

\bibitem{30} Y. Oktem, R. B. Cakirli, R. F. Casten, R. J. Casperson and D. S. Brenner, Phys. Rev. {\bf C74}, 027304 (2006).

\bibitem{31} M. Stoitsov, R. B. Cakirli, R. F. Casten, W. Nazarewicz and W. Satula, Phys. Rev. Lett. {\bf 98}, 132502 (2007).

\bibitem{12} G. Audi, A. H. Wapstra and C. Thibault, Nucl. Phys. {\bf A729}, 337 (2003).

\end{thebibliography}
\end{document}